\documentclass[preprintnumbers, prl, showpacs, floatfix,twocolumn,
preprintnumbers, letterpaper, superscriptaddress]{revtex4}
\usepackage{amsfonts}
\usepackage{amsmath}
\usepackage{amssymb,epsf}
\usepackage{latexsym}
\usepackage{graphicx,epsfig}
\usepackage{amssymb}
\usepackage{subfigure}
\usepackage[colorlinks=true,citecolor=blue,linkcolor=blue,urlcolor=black]{hyperref}
\usepackage{epstopdf}
\usepackage{color}

\def\be{\begin{equation}}
\def\ee{\end{equation}}
\def\ba{\begin{eqnarray}}
\def\ea{\end{eqnarray}}

\def\ra{\rangle}

\begin{document}
\title{Universal representation of the long-range entanglement in the family of Toric Code states}
\author{Mohammad Hossein Zarei}
\email{mzarei92@shirazu.ac.ir}
\affiliation{Physics Department, College of Sciences, Shiraz University, Shiraz 71454, Iran}
\author{Mohsen Rahmani Haghighi}
\email{Rahmani.qit@gmail.com}
\affiliation{Physics Department, College of Sciences, Shiraz University, Shiraz 71454, Iran}
\begin{abstract}
 Since the long range entanglement is a universal characteristic of topological quantum states belonging to the same class, a suitable mathematical representation of the long range entanglement has to be also universal. In this Letter, we introduce such a representation for the family of Toric Code states by using Kitaev's Ladders as building blocks. We consider Toric Code states corresponding to various planar graphs and apply non-local dientanglers to qubits corresponding to non-contractible cycles that satisfy a topological constraint. We demonstrate that, independent of the geometry of the underlying graph, disentanglers convert Toric Code states into a tensor product of Kitaev's Ladder states. Since Kitaev's Ladders with arbitrary geometric configurations include the short-range entanglements, we conclude that the above universal and non-local pattern of entanglement between ladders is responsible of the long-range entanglement inherent in Toric Code states. Our result emphasizes in the capability of such non-local representations to describe topological order in ground-state wave functions of topological quantum systems. 
\end{abstract}
\pacs{03.65.Ud, 03.65.Vf, 64.70.Tg, 02.10.Ox}
\maketitle
\textbf{Introduction}- Various macroscopic phases of matter emerge from different organizations of microscopic building blocks. In symmetry breaking phases \cite{3,2,4}, each organization is described by a specific symmetry and local order parameters can characterize various phases. On the other hand, in topological phases of matter \cite{5,7,8,35}, microscopic building blocks are organized based on topological invariants instead of symmetry \cite{9,10,6,11,gau} and the corresponding order has a non-local nature \cite{12,13,15,16,17}. In quantum many-body systems, entanglement plays a key role in understanding different nature of topological quantum phases. While entanglement in quantum states corresponding to symmetry breaking phases is short-range, topological quantum states have long-range entanglements \cite{18,enta,enta1}. In particular, it is expected that quantum states belonging to various topological classes are characterized by different long-range entanglement patterns \cite{14,to,21,20,23,22}.

On the other hand, it is known that entanglement in a quantum many-body state has a structure which plays key role in describing critical quantum systems \cite{T1}. The entanglement structure in quantum states can be described by mathematical representations such as tensor networks \cite{T2,T3}. Such representations are extensively used in analytical and numerical studies on complex quantum systems \cite{T4,T5,T6,ten}. They are specifically known useful tools for identification of topological order in the ground state wave functions \cite{mat1,mat2,mat3}. An important approach in this direction is entanglement renormalization \cite{24,25,reno,reno2} where short-range entanglements of a topological lattice model is removed along steps of renormalization and the fixed point is a long-range entangled state \cite{30}. However, there is still no representation that exclusively describe the long-range component of the entanglement in a fixed point state. Moreover, while the above known representations are local, it is expected that a non-local representation can be better matched to long-range nature of entanglement in topological quantum phases.
  
Among different topological systems, Toric Code (TC) \cite{6} is a simple and important toy model for studying entanglement structure \cite{Ki1,k3,k6,k7,k8} and it is known as the fixed point of $Z_2$ topological phase under entanglement renormalization \cite{30}. Moreover, recently it has been shown that a TC state on square lattice can be converted to tensor product of Kitaev's Ladder states by applying non-local disentanglers between different layers \cite{26}. It proposes that we can consider one-dimensional Kitaev's Ladders as building blocks of the TC state and look for the long-range entanglement pattern in terms of non-local pattern of entangling ladders. Our main message in this Letter is to show that there is a universal, non-local structure of the entanglement in the family of TC states and it can be regarded as a non-local representation of the long-range entanglement.

 To this end, we notice that TC states can be defined on any arbitrary planar graph and accordingly, we discover an interesting correspondence between a layered connectivity in a planar graph and a layered structure of the entanglement in the corresponding TC state. Such mapping provides a non-local representation for the family of TC states. We show that this representation is universal and independent of the geometry of the underlying graph and therefore, it describes the structure of longe-range entanglemet in TC states. In particular, for different planar graphs we consider non-trivial cycles with a topological constraint where by removing the edges belonging to cycles, the initial graph as well as its dual are converted to one-dimensional disconnected graphs. Then, we show that, by applying suitable non-local disentanglers to qubits corresponding to the above edges of the graph, the initial TC state is converted to tensor product of Kitaev's Ladder states. Since Kitaev's Ladder states are in a topologically trivial phase \cite{33,l1,32,sym}, it is clear that entanglement between ladders is responsible for the long-range entanglement in TC states. It is in fact a description of topological order in terms of topological organization of entanglement between one-dimensional building blocks.
 
\begin{figure}
	\centering
	\includegraphics[width=8cm,height=5cm,angle=0]{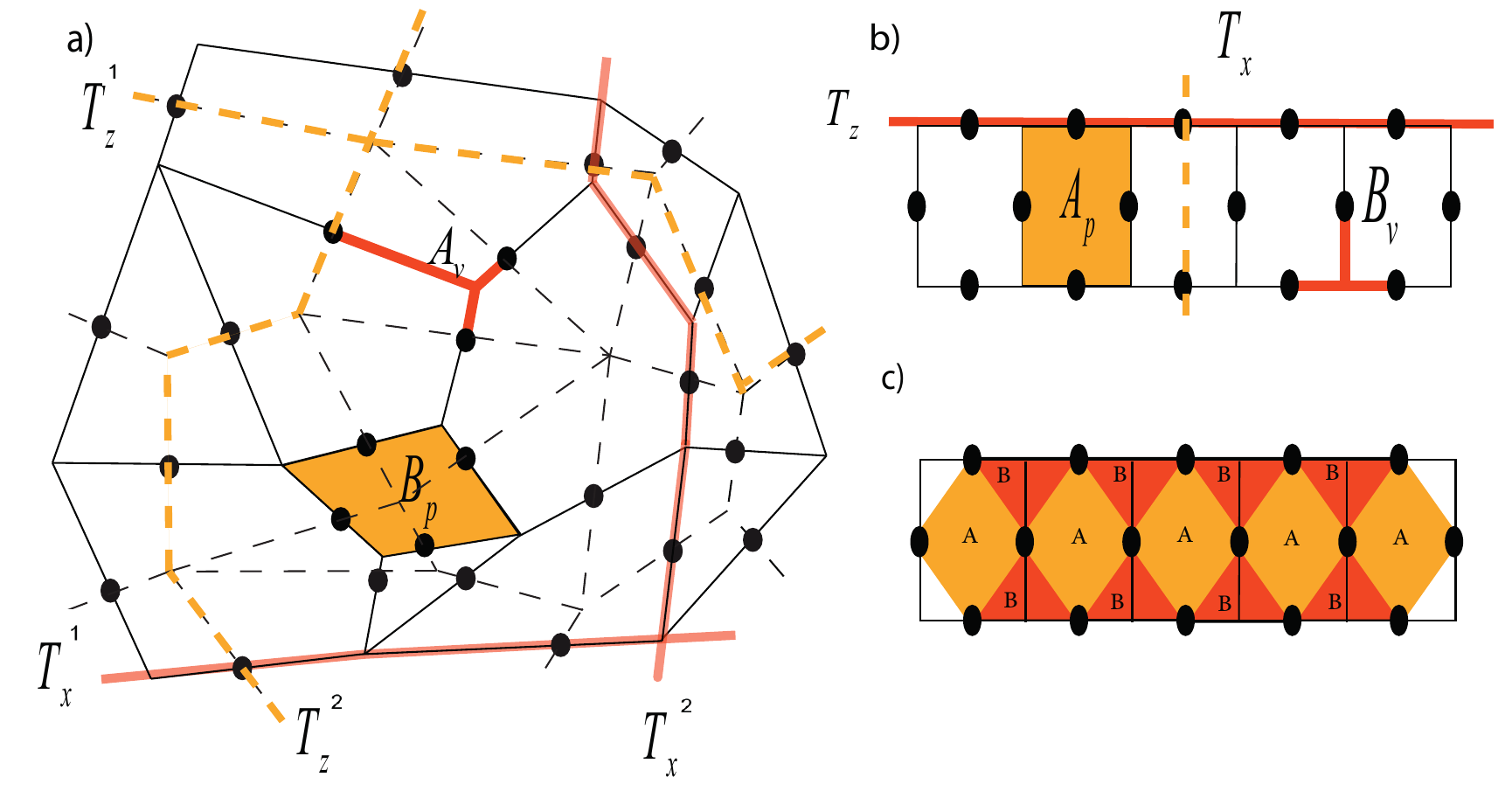}
	\caption{(Color online) a)TC on a planar graph where qubits live on edges. Non-trivial cycle operators of $T_x ^1 $, $T_x ^2$ and $T_z ^1$, $T_z ^2$ are defined on graph (solid line) and dual graph (dashed line) respectively. b) Kitaev's Ladder on square ladder. While $T_z$ is non-local, $T_x$ is a local operator. c) Checkerboard representation of Kitaev's Ladder} \label{planar}
\end{figure}

\textbf{Layered structure of triangular TC}- TC can be defined on arbitrary planar graphs on a torus with qubits living on edges of the graph, see Fig.(\ref{planar}-a). The corresponding Hamiltonian is defined in terms of commutative vertex and plaquette operators $A_v =\prod_{i\in v}Z_i $ and $B_p =\prod_{i\in \partial p} X_i$ in the form of $H_{TC}=-\sum_v A_v -\sum_p B_p$ where $X$ and $Z$ are Pauli operators and $i\in \ v$ refers to qubits that live on edges incoming to a vertex $v$ and $i\in \partial p$ refers to qubits that live on edges around a plaquette $p$. The ground state of the above Hamiltonian has a four-fold degeneracy which is characterized by non-trivial cycle operators in the form of product of $X$ or $Z$ operators around the torus. As shown in Fig.(\ref{planar}-a), we consider non-trivial cycle operators $T_x ^1$ (defined on a cycle of the graph) and $T_z^1$ (dfined on cycle of dual graph) which are commuted with each other and with all $B_p$ and $A_v$ operators. Then, the set of commutative operators of $\{B_p, A_v, T_x ^1 ,T_z ^1 \}$ generate a stabilizer group. One of the ground states of the TC, that we call it $|G_{00}\rangle$, is stabilized by the above set. There are also two other non-trivial cycle operators $T_x ^2$ and $T_z^2$ which anticommute with $T_z ^1$ and $T_x^1$, respectively. Therefore, other ground states are generated by applying $T_x ^2$ and $T_z^2$ on $|G_{00}\rangle$.
\begin{figure*}
	\centering
	\includegraphics[width=17cm,height=5cm,angle=0]{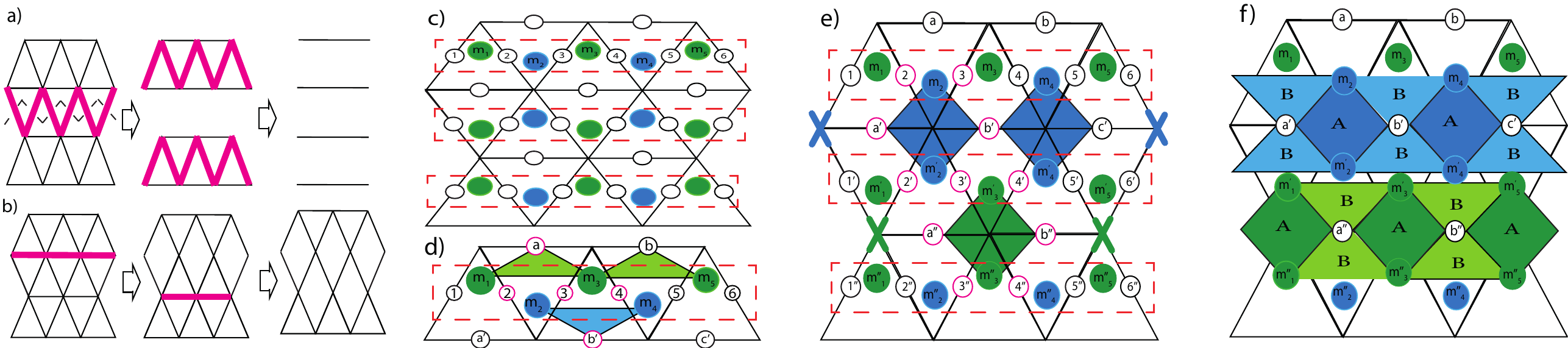}
	\caption{(Color online) a) Topological cycles on a triangular lattice. Removing edges of such cycles leads to disconnected chains. A topological cycle is also a cycle in the dual lattice (dashed line). b) Removing edges of ordinary cycles does not lead to disconnected chains. c) GHZ transformation is applied to qubits belonging to topological cycles. GHZ qubits are denoted by blue and green circles. d,e) GHZ transformation on plaquette and vertex operators generates decoupled stabilizers on blue and green GHZ qubits. f) Initial stabilizers are converted to decoupled Kitaev's Ladders.} \label{ttriangle}
\end{figure*}
 
 One can also define a quasi one-dimensional version of the TC on a ladder which we call Kitaev's Ladder, see Fig.(\ref{planar}-b). Notice that, unlike Toric code, here we define vertex (plaquette) operators as product of $X$'s ($Z$'s)and therefore we denote them by $B_v$ ($A_p$). Stabilizers can also be represented by a checkerboard pattern as shown in Fig.(\ref{planar}-c). There is also a two-fold degeneracy which is characterized by two non-commutative operators $T_z$ and $T_x$ shown in Fig.(\ref{planar}-b). In particular, one of the ground states that we call $|L_0\rangle$ is characterized by set of stabilizers including $A_v$'s $B_p$'s and $T_z$ while another ground state is $T_x |L_0\rangle$.
 
 The Kitaev's ladder is topologically distinct from the TC. Since topological order is a property of the wave function, we expect that the above distinction is realized in terms of different properties of quantum states $|G_{00}\rangle$ and $|L_0\rangle$. In other words, $|G_{00}\rangle$ as a quantum state in the ground state subspace of the toric code should be compared with $|L_0 \rangle$ as a quantum state in the ground state subspace of the Kitaev's ladder model. In particular, while $|G_{00}\rangle$ is a long-range entangled state, $|L_0\rangle$ has a short-range entanglement \cite{33,l1,32}. In the following, we find an interesting connection between these states where we show Kitaev's ladder states are building blocks of the TC state and the long-range entanglement in the TC state is represented by a topological pattern of entanglement between Kitaev's ladders. 
 
 We use idea of layer-by-layer disentangling for square TC introduced in \cite{26}. We start our study by considering the TC on a triangular lattice and then extend it to arbitrary planar graph. As shown in Fig.(\ref{ttriangle}-a,b), there are different types of non-contractible cycles for triangular lattice. However, we consider a specific type of cycles which satisfy a topological constraint. In particular, in graph theory, a topological constraint is defined by three sets of edges of a planar graph namely $C$, $Int_c$ and $Out_c$ \cite{tcycle}. $C$ refers to set of edges belonging to the cycle and $Int_c$ and $Out_c$ refer to sets of edges which are inter or outer of the cycle with a constraint that $Int_c \cap Out_c= \emptyset$ which means $Int_c$ and $Out_c$ do not have any common vertex. As shown in Fig.(\ref{ttriangle}-a), because of the above constraint, by removing edges belonging to a topological cycle, the initial graph is separated to two disconnected graph unlike an ordinary cycle as shown Fig.(\ref{ttriangle}-b). In this regard, if we remove all parallel topological cycles in the triangular lattice, it is separated to several one-dimensional disconnected chains, see Fig.(\ref{ttriangle}-a). We also notice that a topological cycle has an interesting property that it corresponds also to a cycle in the dual graph as seen in Fig.(\ref{ttriangle}-a).

 Now we consider $|G_{00}\rangle$ for the triangular Toric code, and show that corresponding to the above layered structure of triangular lattice, there is a layered structure regarding entanglement in the TC state. Our main idea is to apply unitary transformations as layer-by-layer disentanglers in the sense that $|G_{00}\rangle$ is converted to a tensor product of Kitaev's ladder states.

 In order to define our unitary transformations, we consider chains of qubits including N qubits belonging to the topological cycles, see Fig.(\ref{ttriangle}-c). For each chain, we define an N-qubit GHZ basis which is defined by a set of stabilizers in the form of $g^{(N)}=\lbrace Z_{1} Z_{2}, Z_{2} Z_{3}, ... ,Z_{N}Z_{1}X_{1}X_{2}...X_{N}\rbrace$. Then, by denoting these operators by $g_i$'s, the GHZ basis is defined in the following form:
\begin{equation}\label{new basis}
\prod_{i=1}^{N}|\bar{m_i}\ra=\frac{1}{2^{N/2}}\prod_{i=1}(1+(-1)^{m_i}g_i)|++...++0\ra
\end{equation}
Where $m_i=0,1$ and $|+\rangle$ and $|0\rangle$ are positive eigenstates of the Pauli operators $X$ and $Z$, respectively. One can denote these variables by new qubits which are placed between the two initial qubits i and i+1 on a chain and we call them GHZ qubits. It is clear that there is a non-local unitary operator that converts the computational basis to the above GHZ basis. We call such a unitary transformation a GHZ disentangler. 

In order to find the effect of the above dientanglers to the $|G_{00}\rangle$, we notice that there is a one-to-one correspondence between a stabilizer state and the group of its stabilizers. It means that instead of considering the effect of unitary transformations on $|G_{00}\rangle$, we can consider our transformation on stabilizer operators corresponding to $|G_{00}\rangle$. Then, the transformed quantum state is exclusively determined by considering group of the transformed stabilizers. In this regard, we should apply our unitary transformations to stabilizers of $|G_{00}\rangle$ including operators generated by $A_v$'s, $B_p$'s, $T_z ^1 $ and $T_x ^1$ or equivalently we rewrite these stabilizers in the GHZ basis. Notice that it does not mean that we consider the effect of the GHZ transformation on the Hamiltonian of the toric code. In particular, $T_z ^1$ and $T_x ^1$ are not terms in the Hamiltonian. Our goal is to do transformation on stabilizers of $|G_{00} \rangle$ and then to determine the final transformed state by looking on the new stabilizers.

Since initial stabilizers are in the form of product of $X$ or $Z$ operators, we first introduce two elementary GHZ transformations on Pauli operators as follows:

\textbf{$X$-transformation:} Consider the effect of pauli operator $X_j$ on $j$th qubits, from a GHZ chain. We write $X_j\prod_{i=1}^{N}|\bar{m_i}\ra$ as $ \prod_{i\neq j-1 , j}(1+(-1)^{m_i}g_i)[X_j (1+(-1)^{m_{j-1}}g_{j-1})(1+(-1)^{m_j}g_j)]|++...+0\ra$. Since $g_{j-1}=Z_{j-1}Z_j $ and $g_j =Z_j Z_{j+1}$ unticommute with $X_j$, we conclude that $X_j (1+(-1)^{m_{j-1}}g_{j-1})(1+(-1)^{m_j}g_j)=	(1+(-1)^{m_{j-1}+1}g_{j-1})(1+(-1)^{m_{j} +1}g_j)X_j$. Then, since $X_j |+\rangle =|+\rangle$, for $j\neq N$ we conclude $X_j|\bar{m}_{j-1}\ra |\bar{m}_j\ra =|\bar{m}_{j-1}+1\ra|\bar{m}_j +1\ra$. Therefore, the effect of $X_j$ on physical qubit $j$ plays the role of logical pauli operator $\bar{X}_{j-1}  \bar{X}_j$ on the GHZ qubits of $|\bar{m}_{j-1}\ra |\bar{m}_j\ra$ and we call it an $X$-transformation. 

\textbf{$Z$-transformation:} consider the effect of a product of $Z$ operators on two consecutive qubits of the GHZ chain in the form of $Z_j Z_{j+1}$. Since, for  $j \neq N$, $Z_j Z_{j+1}$ is the same as $g_j$ we have $g_j\prod_{i=1}^{N}|\bar{m}_i\ra= \prod_{i\neq j}(1+(-1)^{m_i}g_i) [g_j (1+(-1)^{m_j}g_j)]|++...+\ra$. Since $g_j ^2 =1$, it is simple to check that $g_j (1+(-1)^{m_j}g_j) =(-1)^{m_j} (1+(-1)^{m_j}g_j)$. Therefore, it is concluded that $g_j|\bar{m}_j\ra=(-1)^{m_j}|\bar{m}_j\ra$ which means that the effect of $g_j =Z_j Z_{j+1}$ on $j$ and $j+1$th physical qubits plays the role of a logical Pauli operator $\bar{Z}_j$ on the $j$th GHZ qubit $|\bar{m}_j\ra$ and we call it a $Z$-transformation.

 Using the above elementary transformations, we are ready to consider the effect of $GHZ$ disentanglers on $A_v$ and $B_p$ stabilizers in the triangular TC: 
 
 1. First we consider transformation of plaquette operators. As shown in Fig.(\ref{ttriangle}-d), consider a plaquette operator of $B_p=X_2 X_3 X_a$. Using the $X$-transformation, $X_2$ and $X_3$ are transformed to $\bar{X_1}\bar{ X_2}$ and $\bar{X_2}\bar{ X_3}$, respectively while $X_a$ remains unchanged. Therefore, $B_p$ is transformed to $X_a \bar{X_1} \bar{X_3}$ that we denote it by $\bar{B}_{\mathfrak{g}}$. In the same way, the next plaquette operator of $B_p=X_3 X_4 X_{b^\prime}$ shown in Fig.(\ref{ttriangle}-d) is transformed to  $X_{b^\prime} \bar{X_2} \bar{X_4}$ that we denote it by $\bar{B}_{\mathfrak{b}}$. Interestingly $\bar{B}_{\mathfrak{g}}$'s and $\bar{B}_{\mathfrak{b}}$'s are completely decoupled because they are applied to different sets of GHZ qubits corresponding to odd and even numbers which are denoted by two different colors in Fig.(\ref{ttriangle}-d). 
 
2. In the next step, we consider vertex operators. As shown in Fig.(\ref{ttriangle}-e), consider a vertex operator of $A_v=Z_{a'} Z_{b'} Z_2 Z_3 Z_{2'} Z_{3'}$.  Using the $Z$-transformations, $Z_2 Z_3$ and $Z_{2'} Z_{3'}$ are transformed to logical operators $\bar{Z}_2$ and $\bar{Z}_{2'}$, respectively. Consequently, the initial operator $A_v$ is transformed to $Z_{a'} Z_{b'} \bar{Z}_2 \bar{Z}_{2'}$ which we denote it by $\bar{A}_{\mathfrak{b}}$. In the same way, another vertex operator of $A_v=Z_{a''} Z_{b''} Z_{3'} Z_{4'} Z_{3''} Z_{4''}$ in the next row of the lattice, is transformed to $Z_{a''} Z_{b''} \bar{Z}_{3'} \bar{Z}_{3''} $ which we denote it by $\bar{A}_{\mathfrak{g}}$. $\bar{A}_{\mathfrak{b}}$'s and $\bar{A}_{\mathfrak{g}}$'s are applied to GHZ qubits with even (blue) and odd (green) numbers, respectively and therefore they are completely decoupled.
\begin{figure*}
	\centering
	\includegraphics[width=18cm,height=7cm,angle=0]{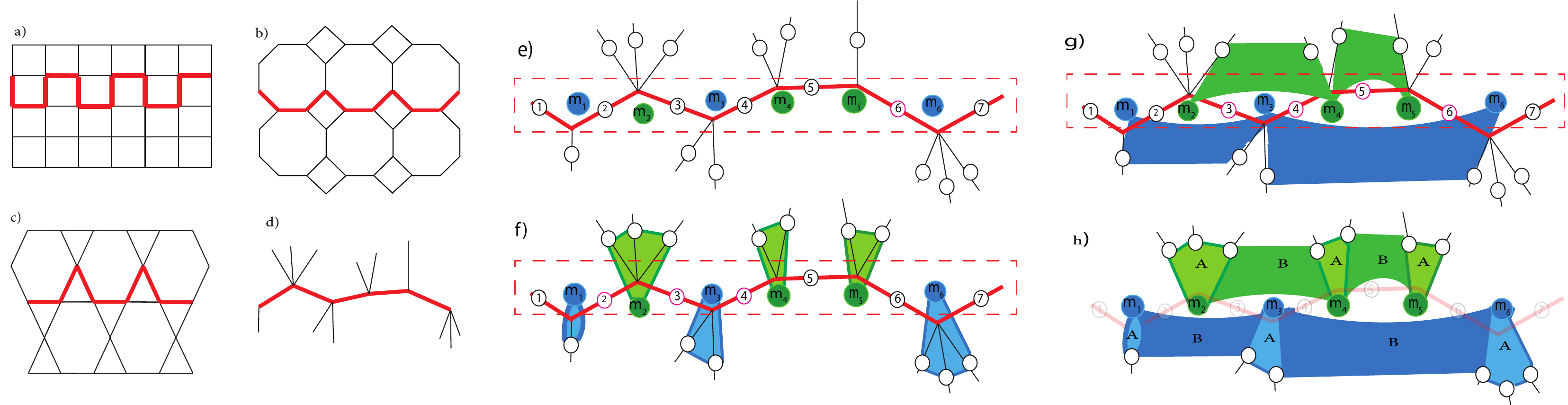}
	\caption{(Color online) a,b,c) Topological cycles on three different lattices d) a general form of a topological cycle where each vertex of the cycle has incoming edges or outgoing edges and no both of them. e) GHZ transformation for an arbitrary topological cycle. f,g) GHZ transformation on vertex and plaquette operators generates decoupled stabilizers on blue and green GHZ qubits. h) Final stabilizers follow checkerboard patterns corresponding to decoupled Kitaev's Ladders. } \label{arbitrary}
\end{figure*}

3. On the other hand, as seen in Eq.\ref{new basis}, the final qubit of the GHZ chain is set initially to $|0\rangle$. Therefore, transformation on stabilizers including $N$th qubit should be carefully considered. However, there is an important point that since the operator $T_x ^1 =X_1 X_2 ... X_N $ is also a stabilizer of the $|G_{00}\rangle$, we can replace each generator of $A_v$ or $B_p$ with $A_v T_x ^1$ or $B_p T_x ^1$. For example, the operator $Z_N Z_1$ is replaced by $Z_N Z_1 T_x ^1$ which is the same as $g_N$ and it is transformed by the $Z$-transformation to $\bar{Z}_N$. The operator $X_N$ is also replaced by $X_N T_x ^1$ which is equal to $X_1 X_2 ... X_{N-1}$ and it is transformed to $\bar{X}_{N-1} \bar{X}_{N}$ by the $X$-transformation. It means that stabilizers corresponding to the boundary qubits are transformed similar to other stabilizers.

We apply the above transformations on all stabilizers of $|G_{00} \rangle$. As shown in Fig.(\ref{ttriangle}-f), $\bar{A}_{\mathfrak{g}}$ and $\bar{B}_{\mathfrak{g}}$ ($\bar{A}_{\mathfrak{b}}$ and $\bar{B}_{\mathfrak{b}}$) operators match to a checkerboard structure similar to the Kitaev's Ladder in Fig.(\ref{planar}-c). There are in fact two decoupled green and blue Kitaev's Ladders. However, it still remains to consider GHZ transformation on non-trivial stabilizers $T_z^1 $ and $T_x ^1$. However, we consider $T_x ^1 T_z^1$ instead of $T_x ^1$. Then, notice that $N$ is an even number because of periodic boundary condition and accordingly $T_z^1 = Z_1 Z_2 ... Z_{N-1}Z_N$  and $T_x ^1 T_z^1= Z_2 ... Z_{N-1}Z_N Z_1 T_x^1$ are equal to $g_1 g_3 ...g_{N-1}$  and $g_2 g_4 ... g_N$, respectively. Then, by transformation of $g_i$ to $\bar{Z}_i$, $T_z ^1$ and $T_x ^1 T_z^1$ are transformed to $\bar{Z}_1...\bar{Z}_{N-1}$ and $\bar{Z}_2...\bar{Z}_{N}$, respectively. These operators are in fact non-trivial stabilizers of $T_z$ for $|L_0\rangle$ on green and blue Kitaev's ladders. We should emphasize in importance of the topological cycle on the above simple transformation of non-trivial operators. The point is that a topological cycle has a property that it is also a cycle in dual lattice. Therefore, we was able to consider both $T_{x}^1$ and $T_{z}^1$ on a topological cycle while $T_z^1$ can not be defined on an ordinary cycle.

  Finally, after applying GHZ disentanglers corresponding to all chains in the triangular lattice, the initial TC state is transformed to a tensor product of Kitaev's Ladder states, $U_{GHZ}|G_{00}\rangle=|L_0\rangle^{\otimes M}$ where $M$ refers to number of layers.

\textbf{Extension to arbitrary planar graphs}- Our main message in this Letter is that the above disentangling pattern is a universal feature of the family of TC state independent of the underlying graph in which TC is defined and therefore it is a non-local representation of the long-range entanglement in the TC state. To this end, here we should consider other planar graphs. The main point is that topological cycles that we introduced for the triangular lattice can also be found for other geometries. In Fig.(\ref{arbitrary}-a,b,c), for example we show such cycles for square, Kagomme and square-octagonal lattices. 

In an arbitrary planar graph, although vertices of the graph along a topological cycle can have different connections, topological constraint forces them to have incoming edges or outgoing edges and not both of them. In Fig.(\ref{arbitrary}-d), we show such an arbitrary topological cycle. Now, we consider the effect of the GHZ transformation corresponding to such a cycle on the corresponding TC. As shown in Fig.(\ref{arbitrary}-e), we consider two different colors for GHZ qubits where GHZ qubits living above the cycle are denoted by green color and those living below the cycle are denoted by blue colors: 

1. We consider GHZ transformation on vertex operators. In Fig.(\ref{arbitrary}-e), consider a vertex living above a green GHZ qubit such as $m_2$. Such a vertex operator have two common qubits 2 and 3 with the cycle. Therefore, using the $Z$-transformation, $Z_2 Z_{3}$ in the corresponding vertex operator is transformed to $\bar{Z}_2$ applied to $m_2$. In the same way for a vertex below a blue GHZ qubit such as $m_3$, $Z_3 Z_4$ is transformed to $\bar{Z}_3$ applied to $m_3$. As shown in Fig.(\ref{arbitrary}-f), this simple argument shows similar to the triangular lattice, here again vertex operators of the TC along the topological cycles are transformed to two decoupled sets of stabilizers corresponding to blue and green GHZ qubits.
 
 2. For plaquette operators, consider a plaquette which includes three qubits 4, 5, 6 shown in Fig.(\ref{arbitrary}-e). We name it a down plaquette because it is below the cycle. As shown in Fig.(\ref{arbitrary}-g), by the $X$-transformation, $X_4 X_5 X_6$ in the corresponding plaquette operator is transformed to $\bar{X}_3 \bar{X}_6$ applied to blue GHZ qubits $m_3$ and $m_6$. On the other hand, for a top plaquette which includes qubits 3 and 4, $X_3 X_4$ in the corresponding plaquette operator is transformed to $\bar{X}_2 \bar{X}_4$ applied to green GHZ qubits of $m_2$ and $m_4$. In this regard, the down and top plaquette operators are transformed to operators defined on blue and green GHZ qubits, respectively. 
 
  As shown in Fig.(\ref{arbitrary}-h), one can see the checkerboard structure of new stabilizers similar to what was derived for the triangular lattice. Moreover, it is simple to check that non-trivial stabilizers of $T_z^1$ and $T_x^1 T_z^1$, defined on the topological cycle, are also converted to two decoupled $T_z$ operators for blue and green GHZ qubits similar to the triangular case. In this regard, by applying GHZ transformations to all parallel topological cycles in the initial planar graph, we would have a tensor product of Kitaev's Ladder states which live between topological cycles. We notice that while geometry of final ladders depends on the geometry of connections of vertices in topological cycle, topological constraint is only important parameter for decoupling in a checkerboard pattern. On the other hand, since Kitaev's Ladder states have no topological order, different geometries are not important in pattern of long-range entanglement. Therefore, we represent the long-range entanglement in the $|G_{00}\rangle$ by pattern of entanglement between ladders which is universal, non-local and is based on a topological constraint.
 
\begin{figure}
	\centering
	\includegraphics[width=6cm,height=4cm,angle=0]{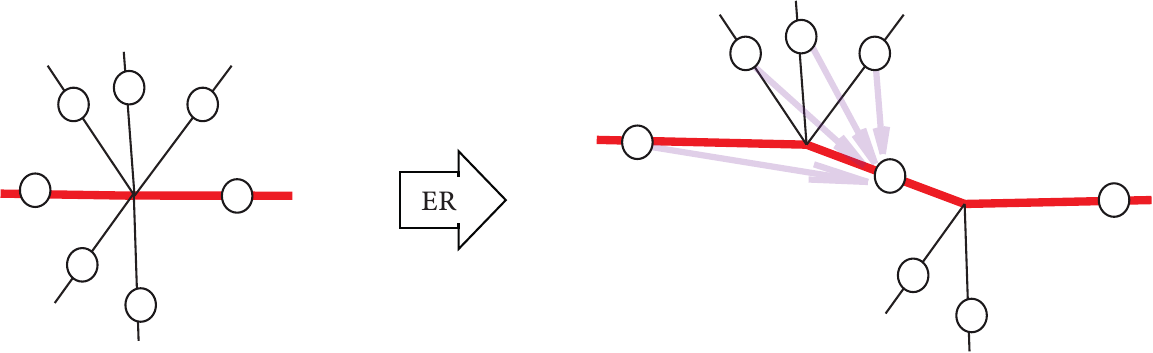}
	\caption{(Color online) Red line is a part of a cycle. An ordinary cycle is transformed to a topological cycle under entanglement renormalization. Arrows in the right hand represent CNOT operators.} \label{ER}
\end{figure}

Finally, we emphasize that it was very important in our disentangling mechanism that both non-trivial operators $T_z^1$ and $T_x^1$ could be defined on the topological cycles. However, it is not clear that for arbitrary planar graph there are topological cycles or not. It is important for proving our claim that the layered entanglement structure is a universal property of the TC state. To this end, we use a simple local transformation regarding to entanglement renormalization in the TC presented in \cite{30}. As shown in Fig.(\ref{ER}), consider a part of a non-topological cycle where a vertex has both incoming and outgoing edges. However, as shown in \cite{30}, such a vertex can split to two vertices by adding a new edge qubits (isometric transformation)  and then applying CNOT operators according to local pattern shown in the figure. Therefore, after entanglement renormalization we have a topological cycle where a vertex has incoming edges and another vertex has outgoing edges.

\textbf{Conclusion}- Our study shows that the long-range entanglement in the TC state is well described by a non-local representation in real space. Because of non-local nature of topological order, it is anticipated that such non-local representations of quantum states will be better matched with topological quantum systems compared to known local representations. Furthermore, real space representations offer the advantage of naturally describing the entanglement structure through the geometric configuration of the underlying lattices in topological lattice models. In this regard, it would be important to consider similar non-local representations for other topological systems. In particular, representations with distinct topological constraints are expected to govern quantum states associated with differing topological classes. This implies that topological constraints can be interpreted as non-local order parameters. Accordingly, non-local representations offer advantages in classifying quantum topological phases. It is specifically an important task for future works to consider how such a representation can be useful for calculating topological entanglement entropy as an important measure of topological order. Finally, we emphasize that mathematical representations of quantum states can serve as variational wave functions for complex quantum Hamiltonians. However, non-local representations such as one proposed here are superior candidates for variational wave functions in topological quantum Hamiltonians where long-range entanglement plays a central role. 

\textbf{Acknowledgment}- We would like to thank A. Ramezanpour for the useful discussions which led to the improvement of our argument in the Letter. 

\end{document}